\documentclass[aps,prb,amsmath,amssymb,10pt,preprint,superscriptaddress]{revtex4-1}
\usepackage[utf8x]{inputenc}
\usepackage[english]{babel}
\usepackage{amsmath}
\usepackage{amssymb}

\usepackage{amstext}
\usepackage{graphicx}
\usepackage{lmodern}

\usepackage{float}
\usepackage{siunitx}
\usepackage{color}

\usepackage{titlesec}

\definecolor{MSBlue}{rgb}{.204,.353,.541}
\definecolor{MSLightBlue}{rgb}{.31,.506,.741}
\definecolor{MSRed}{rgb}{.584,0,.118}

\titleformat*{\section}{\large\bfseries\sffamily\color{black}}
\titleformat*{\subsection}{\normalsize\bfseries\sffamily\color{black}}

\titlespacing*{\section}
{0pt}{12 pt}{12 pt}
\titlespacing*{\subsection}
{0pt}{6pt}{6pt}

\usepackage[colorlinks=true]{hyperref}

\newcommand{\antipara}{\mathbin\uparrow\downarrow}
\newcommand{\para}{\mathbin\uparrow\uparrow}

\begin{document}

\title{Ferroic collinear multilayer magnon spin valve}

\author{Joel Cramer}
\affiliation{Institute of Physics, Johannes Gutenberg-University Mainz, 55099 Mainz, Germany}
\affiliation{Graduate School of Excellence Materials Science in Mainz, 55128 Mainz, Germany}

\author{Felix Fuhrmann}
\affiliation{Institute of Physics, Johannes Gutenberg-University Mainz, 55099 Mainz, Germany}

\author{Ulrike Ritzmann}
\affiliation{Department of Physics, University of Konstanz, 78457 Konstanz, Germany}
\affiliation{Institute of Physics, Johannes Gutenberg-University Mainz, 55099 Mainz, Germany}

\author{Vanessa Gall}
\affiliation{Department of Physics, University of Konstanz, 78457 Konstanz, Germany}

\author{Tomohiko Niizeki}
\affiliation{WPI Advanced Institute for Materials Research, Tohoku University, Sendai 980-8577, Japan}

\author{Rafael Ramos}
\affiliation{WPI Advanced Institute for Materials Research, Tohoku University, Sendai 980-8577, Japan}

\author{Zhiyong Qiu}
\affiliation{WPI Advanced Institute for Materials Research, Tohoku University, Sendai 980-8577, Japan}

\author{Dazhi Hou}
\affiliation{WPI Advanced Institute for Materials Research, Tohoku University, Sendai 980-8577, Japan}

\author{Takashi Kikkawa}
\affiliation{WPI Advanced Institute for Materials Research, Tohoku University, Sendai 980-8577, Japan}
\affiliation{Institute for Materials Research, Tohoku University, Sendai 980-8577, Japan}

\author{Jairo Sinova}
\affiliation{Institute of Physics, Johannes Gutenberg-University Mainz, 55099 Mainz, Germany}

\author{Ulrich Nowak}
\affiliation{Department of Physics, University of Konstanz, 78457 Konstanz, Germany}

\author{Eiji Saitoh}
\affiliation{WPI Advanced Institute for Materials Research, Tohoku University, Sendai 980-8577, Japan}
\affiliation{Institute for Materials Research, Tohoku University, Sendai 980-8577, Japan}
\affiliation{Center for Spintronics Research Network, Tohoku University, Sendai 980-8577, Japan}
\affiliation{Advanced Science Research Center, Japan Atomic Energy Agency, Tokai 319-1195, Japan}

\author{Mathias Kläui}
\email{Klaeui@uni-mainz.de}
\affiliation{Institute of Physics, Johannes Gutenberg-University Mainz, 55099 Mainz, Germany}
\affiliation{Graduate School of Excellence Materials Science in Mainz, 55128 Mainz, Germany}

\date{\today}

\begin{abstract}
	Information transport and processing by pure magnonic spin currents in insulators is a promising alternative to conventional charge-current driven spintronic devices. 
	The absence of Joule heating as well as the reduced spin wave damping in insulating ferromagnets has been suggested to enable the implementation of efficient logic devices.
	After the proof of concept for a logic majority gate based on the superposition of spin waves has been successfully demonstrated,
	further components are required to perform complex logic operations.
	A key component is a switch that corresponds to a conventional magnetoresistive spin valve.
	Here, we report on magnetization orientation dependent spin signal detection in collinear magnetic multilayers with spin transport by magnonic spin currents.
	We find in Y\textsubscript{3}Fe\textsubscript{5}O\textsubscript{12}$|$CoO$|$Co tri-layers that the detected spin signal depends on the relative alignment of Y\textsubscript{3}Fe\textsubscript{5}O\textsubscript{12} and Co.
	This demonstrates a spin valve behavior with an effect amplitude of \SI{120}{\percent} in our systems.
	We demonstrate the reliability of the effect and investigate the origin by both temperature and power dependent measurements, showing that spin rectification effects and a magnetic layer alignment dependent spin transport effect result in the measured signal.
\end{abstract}
	
\maketitle

\subsection*{Introduction}

Almost three decades ago, the discovery of the giant magnetoresistance effect (GMR)\cite{baibich1988giant,binasch1989enhanced} led to a key for the research field of spintronics based on spin-polarized charge currents.
The inclusion of the spin degree of freedom in information technology promises, amongst others, the implementation of logic devices with increased speed or capacity as compared to conventional CMOS electronics\cite{wolf2001spintronics}.
This approach has been driven to the next level by the field of magnon spintronics\cite{Chumak2015}.
Magnons are quasiparticles (quanta) of the collective excitation of electron spins (spin waves) in magnetically ordered systems.
As being spin-1 particles, the entity of a great number of magnons with equal propagation direction yields a pure spin current carrying information in the form of angular momentum.
One of the advantages of spin wave mediated information transport, especially in the case of insulating magnetic oxides, is the absence of electron motion and thus the absence of power dissipation due to Joule heating.

Consequently, in view of not only new physical phenomena but also considerable application potential, the investigation of generation and detection of pure magnonic spin currents in insulators attracted significant attention in recent years.
One of the most prominent systems for such spin currents is the insulating ferrimagnet Y\textsubscript{3}Fe\textsubscript{5}O\textsubscript{12} (YIG).
Inherently, YIG exhibits the lowest measured Gilbert damping constant $\alpha \propto 10^{-5}$, enabling long distance spin propagation and thus efficient transport of spin information\cite{Cornelissen2016,Guo2016}.
The detection of magnonic spin currents is typically achieved by means of the inverse spin Hall effect (ISHE)\cite{Sinova2015} and for the generation a number of approaches have been developed: ferromagnetic resonance (FMR) spin pumping\cite{Tserkovnyak2002,Mizukami2002,Azevedo2005,saitoh2006}, thermal generation of spin currents induced by the spin Seebeck effect (SSE)\cite{Uchida2016,Bauer2012,Kehlberger2015}, and the electrical injection of magnons by the spin Hall effect\cite{Cornelissen2015,Goennenwein2015}.

More recently, it was shown that magnon-based logic operations can be realized in structures employing YIG as a spin conduit\cite{Klingler2014,Klingler2015,Fischer2017,Ganzhorn2016}.
The superposition of coherent or incoherent magnons in a spin wave bus or a non-local geometry, respectively, enables the implementation of a fully functional logic majority gate.
Depending on the system, either the phase or the amplitude of the output signal reflects the input majority.
These concrete demonstrations of magnon logic highlight the potential of this new information processing scheme, implying the necessity for further logic building blocks to accomplish more complex operations.
Bearing in mind that contemporary CMOS technology relies on the transistor as the basic unit not just to amplify but in particular as a switch, magnon based components exhibiting similar switching functionality are desirable to allow for a facile transfer of already existing concepts to magnon-based computing.

\subsection*{Experiment}

\begin{figure*}[!bt]
	\centering
	\includegraphics[width = \textwidth]{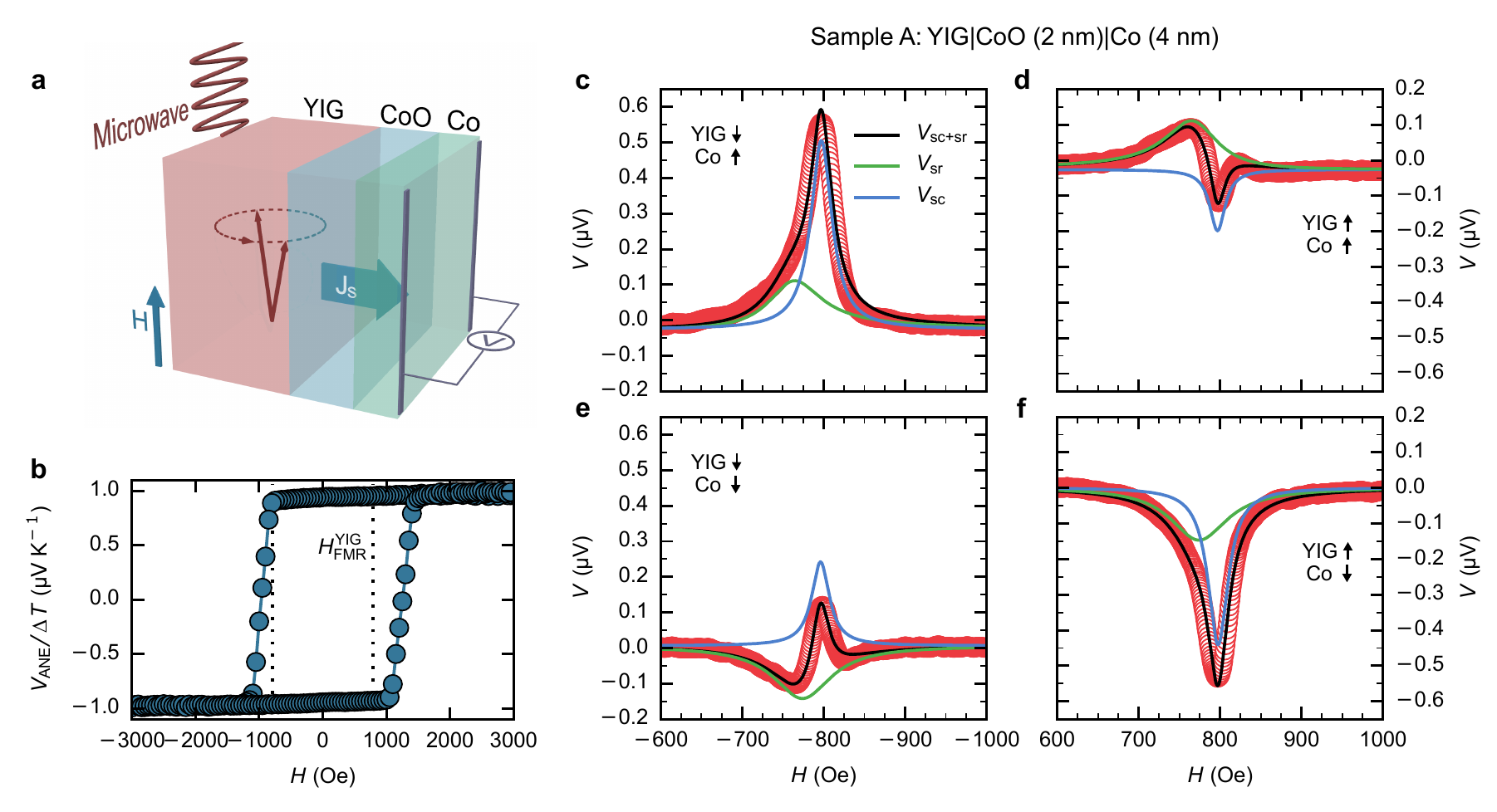}
	\caption{(a) Illustration of the investigated YIG$|$CoO$|$Co tri-layer system.
		A spin current $J_s$ induced in the YIG by FMR spin pumping propagates through the CoO intermediate layer and is detected electrically in the Co layer via the ISHE.
		The relative magnetization orientation of YIG and Co can be adjusted to be either parallel or antiparallel.
		(b) Typical ANE hysteresis loop observed for sample A at $T = \SI{120}{\kelvin}$, being field cooled at $H_{\mathrm{ext}} = \SI{90}{\kilo Oe}$.
		Coercive fields of $H_c^+ \, [H_c^-] \simeq \SI{1250}{Oe} \, [\SI{-970}{Oe}]$ and an elevated squareness $V_{\mathrm{ANE}}^{\mathrm{H=0}}/V_{\mathrm{ANE}}^{\mathrm{sat}} \simeq 0.95$ are observed.
		(d)-(f) Field-dependent voltage signals detected in sample A induced by microwave irradiation ($f = \SI{4.5}{\giga\hertz}$, $P_{\mathrm{abs}}\approx\SI{48}{\milli\watt}$, and $T = \SI{120}{\kelvin}$).
		The total signal is given by a superposition of two distinct signals $V_{\mathrm{sc}}$ and $V_{\mathrm{sr}}$, the sign of which individually depend on the YIG respectively the Co orientation.
		The amplitude of $V_{\mathrm{sc}}$ furthermore depends on the relative alignment of YIG and Co.
		Microwave absorption spectra (see Supplementary Information\cite{Supplemental}) show that FMR of YIG is excited at this frequency at $H_{\mathrm{FMR}}^{\mathrm{YIG}} = \pm \SI{788}{Oe}$.
	}
	\label{fig:figure1}
\end{figure*}

Based on the magnetoresistive spin valve concept, we exploit a magnon spin valve effect by means of magnetization orientation-dependent spin transmission and detection in collinear YIG$|$CoO$|$Co tri-layers.
The sample layout is illustrated in Fig. \ref{fig:figure1}a.
In this system, YIG serves as a spin current conduit and source driven by FMR spin pumping.
The Co layer is used as an active switchable layer as well as for the subsequent spin current detection.
As previously shown, ferromagnets as well exhibit an inverse spin Hall effect (ISHE) allowing for the efficient conversion of a spin current into a charge current\cite{Miao2013,Tian2015}.
The insulating CoO interlayer was chosen as it decouples the ferromagnetic layers while simultaneously allowing for transport of the spin current\cite{Qiu2016}.
Furthermore, CoO becomes antiferromagnetic at low temperatures and thus introduces exchange biasing of the Co film.
Besides a horizontal shift of the Co hysteresis loop, this unidirectional anisotropy yields an enhanced coercive field at which the Co magnetization switches.
The latter allows one to perform spin pumping measurements with parallel and antiparallel alignment of YIG and Co at the FMR resonance field.
For a good signal-to-noise ratio $H_{\mathrm{FMR}}^{\mathrm{YIG}} \gtrsim \SI{800}{Oe}$ is necessary, and such a high switching field is unattainable for thin Co films without exchange biasing.

Both the CoO and the Co layer thickness were varied and here we concentrate on the results of 3 typical stacks.
For reasons of simplification, in the following the investigated trilayers YIG (\SI{5}{\micro\meter})$|$CoO (\SI{2}{\nano\meter})$|$Co (\SI{4}{\nano\meter}), YIG (\SI{5}{\micro\meter})$|$CoO (\SI{3}{\nano\meter})$|$Co (\SI{4}{\nano\meter}), and YIG (\SI{5}{\micro\meter})$|$CoO (\SI{5}{\nano\meter})$|$Co (\SI{6}{\nano\meter}) will be referred to as sample A, B and C, respectively.

\subsection*{Spin-thermoelectric measurements}

To check if the magnetic switching fields of Co are sufficiently high to carry out FMR spin pumping with a good signal-to-noise ratio ($H_{\mathrm{FMR}}^{\mathrm{YIG}} \gtrsim \SI{800}{Oe}$), these are determined by temperature-dependent magneto-galvanic measurements.
The anomalous Nernst effect (ANE) occurring in the Co layer yields a signal proportional to the in-plane magnetization component and thus we can detect the Co switching.
Common magnetometry (e.g. SQUID) yields the YIG switching due to the much smaller volume of the Co film as compared to the YIG layer.
A typical ANE hysteresis curve for sample A recorded at $T$ = \SI{120}{\kelvin} is shown in Fig. \ref{fig:figure1}b.
The graph reveals coercive fields of $H_c^+ \, [H_c^-] \simeq \SI{1250}{Oe} \, [\SI{-970}{Oe}]$ and large squareness with $V_{\mathrm{ANE}}^{\mathrm{H=0}}/V_{\mathrm{ANE}}^{\mathrm{sat}} \simeq 0.95$, implying that Co switches without extensive formation of magnetic domains.
Note that in contrast to previous reports\cite{Lin2016}, the thermal spin current of the YIG film generated by the spin Seebeck effect\cite{Uchida2016,Hou2016} was not observed within the experimental resolution.
A potential explanation for this may be given by different CoO thickness and structures grown on different YIG films (polycrystalline vs. single crystalline)\cite{Prakash2016} and the smaller ISHE in Co compared to the previously used Pt.

\subsection*{Spin signal in YIG$|$CoO$|$Co excited by spin pumping}

Having established the switching properties of both YIG and Co, we study next the spin signal as a function of alignment of the YIG and Co layer.
Fig. \ref{fig:figure1}c-f shows field-dependent voltage signals generated in sample A at $T = \SI{120}{\kelvin}$ induced by $f = \SI{4.5}{\giga\hertz}$ microwave irradiation.
By the application of specific field sweep sequences, parallel (d, e) as well as antiparallel (c, f) alignment of YIG and Co is realized.
In the parallel state (Fig. \ref{fig:figure1}d,e) a multi-peak voltage signal appears.
It is well fitted by two overlapping Lorentzian line shapes of opposite sign, slightly shifted peak field values, and different line widths.
The antiparallel state (Fig. \ref{fig:figure1}c,f), on the other hand, exhibits a voltage peak of one polarity but with a significant asymmetry, which again can be fitted by two overlapping Lorentzian functions.
The comparison of all four datasets allows us to both separate and attribute the peaks to different effects.
Depending on the individual orientation of YIG and Co, the peaks change sign separately and thus can be identified as (i) the signal of the spin current generated by spin pumping from the YIG, transmitted across the CoO and detected by the ISHE in the Co (blue curves in Fig. \ref{fig:figure1}c-f) and (ii) a second signal that depends only on the Co layer direction and thus originates from the Co (green curves in Fig. \ref{fig:figure1}c-f).
We start by exploring the latter for which possible explanations are a thermally induced ANE signal due to dynamic microwave heating\cite{Bakker2012} or an anomalous Hall effect (AHE) induced spin rectification (SR) signal\cite{Gui2007,Azevedo2011,Chen2013,Tee2017}.
The microwave driving current $I_{\mathrm{rf}}$ partially flows through the Co layer by means of capacitive coupling and, together with the out-of-plane component of the Co magnetization precession excited by dipolar coupled YIG magnon modes\cite{Tee2017}, yields a dc voltage.
Since Co is driven off-resonance, the SR signal is expected to be of symmetric shape\cite{Chen2013}, in agreement with the experimental observations.
While both effects can contribute, temperature-dependent measurements support the spin rectification mechanism as discussed below.
As this signal depends only on the Co layer direction, it can thus be easily distinguished from the spin current (SC) transport signal, which is the main thrust of this work.
In the following we denote the signal that depends on the Co orientation as $V_{\mathrm{sr}}$ (green) while the signal that depends on the relative orientation of the two layers is denoted as $V_{\mathrm{sc}}$ (blue).

While the Co layer-dependent effect is based on known mechanisms, the intriguing discovery in this experiment is the alignment-dependent amplitude of $V_{\mathrm{sc}}$ analogous to the spin-polarized charge current transmission in a conventional spin valve.
Whereas $V_{\mathrm{sr}}$ is not depending on the relative alignment of the layers, the amplitude of $V_{\mathrm{sc}}$ is nearly twice as large in the antiparallel alignment state as compared to the parallel state ($V_{\mathrm{sc}}^{\antipara} > V_{\mathrm{sc}}^{\para}$).
Note that qualitatively the same behavior is observed for sample B and C (see Supplementary Information\cite{Supplemental}).

\subsection*{Origin of the spin signals in the YIG$|$CoO$|$Co magnon spin valve}

\begin{figure}[!b]
	\centering
	\includegraphics[width = 8.38 cm]{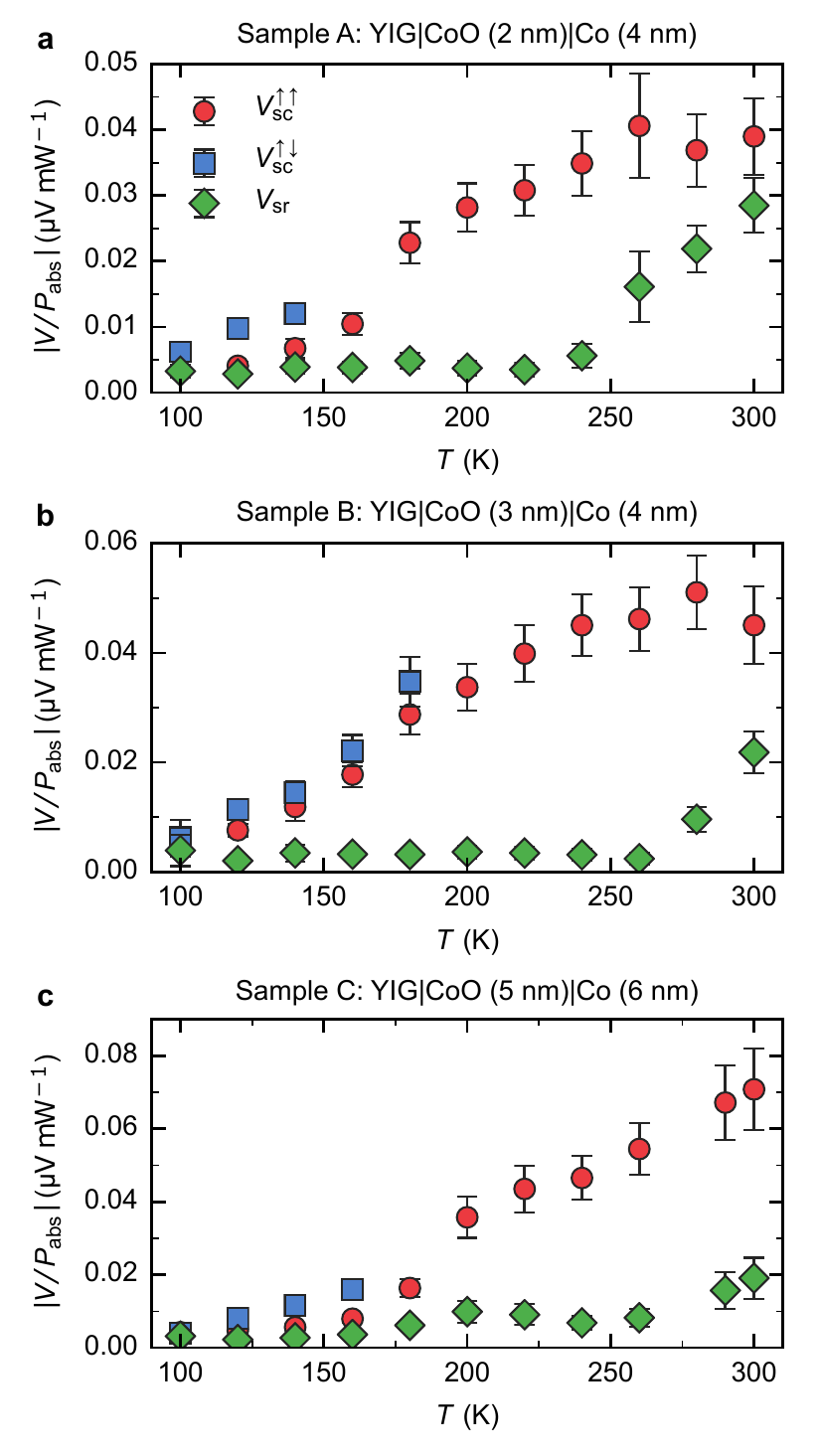}
	\caption{Amplitudes of $V_{\mathrm{sc}}$ for parallel and antiparallel alignment and $V_{\mathrm{sr}}$ normalized by the absorbed microwave power as a function of temperature for (a) sample A, (b) sample B, and (c) sample C.
		Antiparallel alignment of YIG and Co is observed until a critical, sample-dependent temperature.
		The temperature range from \SIrange{100}{300}{\kelvin} was chosen as below \SI{100}{\kelvin} a decreased signal-to-noise ratio impedes straightforward data analysis, whereas higher temperatures were avoided to prevent possible undesirable degradation of the multilayer stack.}
	\label{fig:tempdep}
\end{figure}

To investigate the origin of both signals, temperature-dependent spin pumping measurements were performed, see Fig. \ref{fig:tempdep}a-c.
The spin current transport signal $V_{\mathrm{sc}}$ can be measured up to a critical temperature for both the parallel and the antiparallel state.
This temperature depends on the investigated stack and is due to the fact that above a certain temperature $H_{\mathrm{FMR}}^{\mathrm{YIG}}$ exceeds the coercive field of the Co layer, preventing FMR for antiparallel alignment of the layers.
For the accessible temperature range with antiparallel alignment, we always find $V_{\mathrm{sc}}^{\uparrow\downarrow} > V_{\mathrm{sc}}^{\uparrow\uparrow}$.

In general, the amplitude of $V_{\mathrm{sc}}$ increases with increasing temperature.
Furthermore, in the case of samples A and B a signal maximum is seen at temperatures between \SI{250}{\kelvin} and \SI{300}{\kelvin}.
This behavior can be explained by an enhanced spin conductivity of the CoO layer near its antiferromagnetic-paramagnetic phase transition\cite{Qiu2016}.
It was previously shown that in thin CoO films $T_{\mathrm{N\acute{e}el}}$ is below its bulk value $T_{\mathrm{N\acute{e}el}}^{\mathrm{bulk}} = \SI{290}{\kelvin}$ due to finite size effects and decreases with decreasing film thickness\cite{Ambrose1996,nogues1999exchange}.
Here, temperature-dependent ANE measurements for sample A (see Supplementary Information\cite{Supplemental}) reveal an exchange bias blocking temperature of $T_B \approx \SI{250}{\kelvin}$, suggesting $\SI{250}{\kelvin} \leq T_{\mathrm{N\acute{e}el}} \leq \SI{290}{\kelvin}$\cite{nogues1999exchange} for this specific stack.
In the other multilayers with thicker CoO, $T_{\mathrm{N\acute{e}el}}$ is higher, which explains the monotonic increase of $V_{\mathrm{sc}}$ for $d_{\mathrm{CoO}} = \SI{5}{\nano\meter}$ up to \SI{300}{\kelvin}.

For the amplitude of $V_{\mathrm{sr}}$, the observed temperature dependence is qualitatively different.
Initially, with increasing temperature $V_{\mathrm{sr}}$ remains constant until it starts to increase significantly above a specific temperature, which is characteristic for each stack.
Based on comparison with the Co switching we see that for sample A this temperature coincides with the blocking temperature $T_B$, above which the anisotropy introduced by exchange biasing vanishes.
With regard to the possible mechanism of spin rectification, the disappearance of this additional anisotropy entails a weakened magnetization precession damping.
The magnetization consequently precesses at a larger cone angle\cite{Guan2007}, yielding a larger AHE voltage.
The temperature, above which the amplitude of $V_{\mathrm{sr}}$ starts to increase, is higher for thicker CoO as expected for a thickness-dependent phase transition temperature.
Finally, the observed temperature dependence supports the conclusion that $V_{\mathrm{sr}}$ is not dominated by a magneto-galvanic effect as one would not expect a sudden onset of the signal above \SI{250}{\kelvin} for the ANE.

\begin{figure*}[!b]
	\centering
	\includegraphics[width = 16.38 cm]{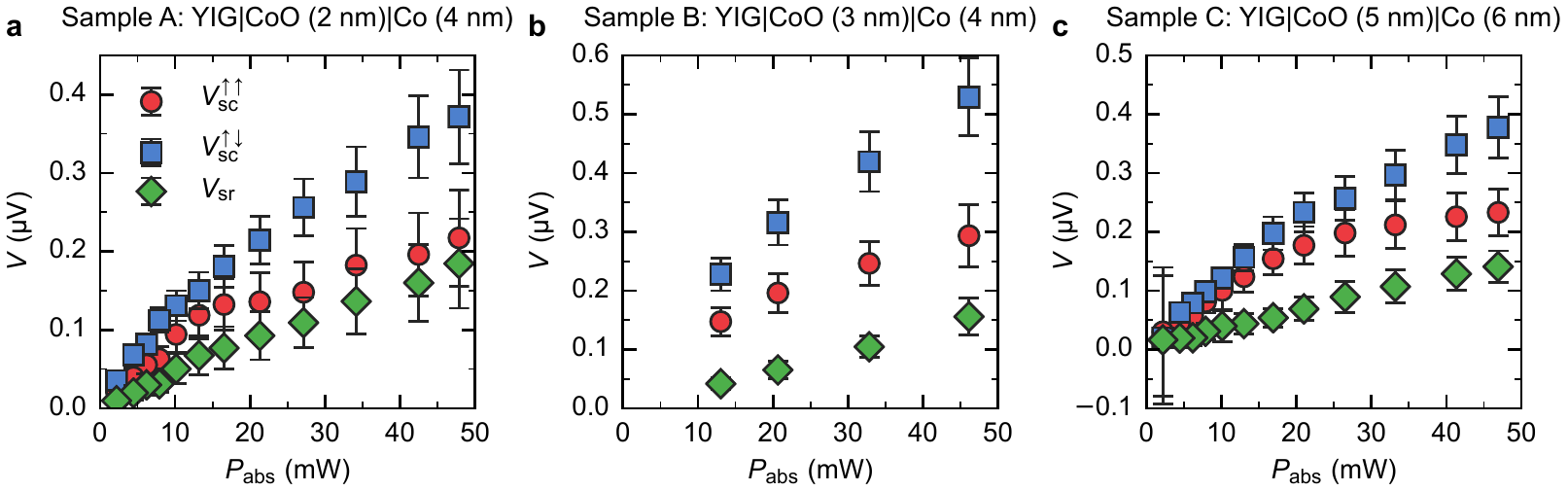}
	\caption{Spin pumping and spin rectification amplitude as a function of the absorbed microwave power for (a) sample A, (b) sample B, and (c) sample C.
		Whereas the amplitude of $V_{\mathrm{sr}}$ exhibits a rather linear power dependence, the one of $V_{\mathrm{sc}}$ follows a non-linear power dependence, indicating an incipient saturation.
		The frequency of the applied microwave is $f = \SI{4.5}{\giga\hertz}$, the system temperature is $T = \SI{120}{\kelvin}$.}
	\label{fig:powdep}
\end{figure*}

Beyond the temperature-dependent data, further information about the signal origin can be obtained from the applied microwave power dependence.
In Fig. \ref{fig:powdep}a-c we show the voltage amplitudes of $V_{\mathrm{sc}}$ and $V_{\mathrm{sr}}$ as a function of the absorbed microwave power.
The spin current transport amplitude exhibits a nonlinear power dependence, regardless of the alignment state.
The onset of a saturation effect is indicated at elevated microwave powers, a behavior that was observed before for spin pumping in YIG/Pt structures\cite{Jungfleisch2015} showing that this signal scales with the generated spin current strength in the YIG.
The power dependence of $V_{\mathrm{sr}}$, on the other hand, does not allow for an equally simple interpretation.
According to Azevedo \textit{et al.}\cite{Azevedo2011} one expects a linear power dependence for $V_{\mathrm{sr}}$ due to its proportionality to the dynamic field driving the Co precession.
While the power dependence observed is in line with this explanation, other possible mechanisms that yield a linear power dependence cannot be excluded.

Finally, we note that the fact that the peak fields for the two effects are slightly different is expected from their different origins.
The peak of the spin rectification signal will be at the field of the maximum of the excited magnon modes in the \SI{5}{\micro\meter} thick YIG crystal that couple to the Co magnetization via dipolar exchange\cite{Tee2017}.
The peak of the spin current transmission signal, however, will be at the field that corresponds to the maximum generation of spin current pumped into the CoO at the YIG|CoO interface.
Since the latter depends on the resonance at the interface while the former covers more of the bulk volume, the observed slight difference is not surprising.

\subsection*{Magnon spin valve effect}

\begin{figure}[!b]
	\centering
	\includegraphics[width = 8.11 cm]{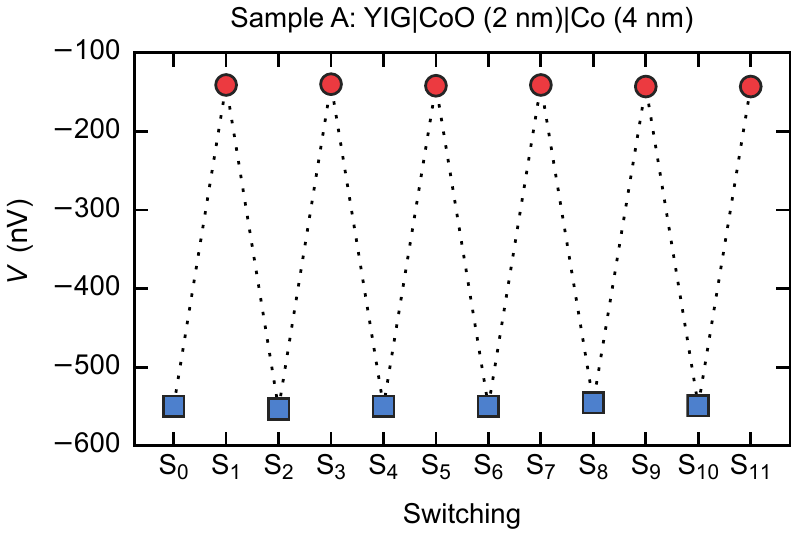}
	\caption{Total voltage for parallel (red circles) and antiparallel (blue squares) alignment of YIG and Co measured in sample A at ferromagnetic resonance $H_{\mathrm{FMR}}^{\mathrm{YIG}} = \SI{788}{Oe}$.
		The absolute voltage difference is \SI{408}{\nano\volt} and the relative change amounts to \SI{290}{\percent}.}
	\label{fig:switch}
\end{figure}

The alignment-dependent spin current transport signal amplitude now naturally lends itself to the implementation of a magnon spin valve effect.
Comparing the amplitude of $V_{sc}$ for parallel and antiparallel alignment in Fig. \ref{fig:figure1}, we find an amplitude of the magnon spin valve effect of $\left( V_{\mathrm{sc}}^{\antipara} - V_{\mathrm{sc}}^{\para} \right) / V_{\mathrm{sc}}^{\para} = \SI{120}{\percent}$ for sample A.
To check the reliability of the magnon spin valve effect, the magnetization direction of the Co top layer is switched several times in a row and the voltage response towards the applied microwave is recorded, see Fig. \ref{fig:switch}.
Since in typical application schemes the external field is fixed instead of being swept and thus one voltage level is probed instead of acquiring the whole absorption spectrum, we choose a single fixed field value at which we acquire the signal amplitude.
We find a large difference in the signal difference of the total voltage for parallel and antiparallel alignment.
For sample A, at $H\approx H_{\mathrm{FMR}}^{\mathrm{YIG}}$ an absolute voltage difference of \SI{408}{\nano\volt} and a spin valve effect amplitude of $\left( V_{\mathrm{sc+sr}}^{\antipara} - V_{\mathrm{sc+sr}}^{\para} \right) / V_{\mathrm{sc+sr}}^{\para} = \SI{290}{\percent}$ is found.
The presented data clearly demonstrates that switching the Co layer yields reliably different and reproducible voltage levels with little scatter.
No distinguishable reduction of the voltage difference was observed for days of measurements of switching, which signifies reliable long term operation without having to reset the magnetization.
Furthermore, the size of the signal change is only weakly temperature dependent so that it will readily also be visible at room temperature for appropriately designed samples with both parallel and antiparallel alignment at the desired resonance field.

\subsection*{Discussion}

To understand our findings of a sizable magnon spin valve effect we have to consider spin current transport and spin-dependent conversion effects. 
Focusing first on the magnonic spin current transport, we note that magnons which are excited in the YIG can enter and travel across the insulating antiferromagnet.
In very thin antiferromagnets this is even true for evanescent modes (see Supplemental Material, where relevant spin model simulations of an FM/AFM/FM trilayer are presented).
The magnons from the YIG layer enter the CoO for both orientations of the YIG with equal probability (note that the CoO does not change its orientation in our experiments as confirmed by probing opposite field cooling directions).
The dispersion relation of the AFM has two branches with opposite sign, such that magnons for both orientations of the YIG are transmitted equally.
In ferromagnets, however, magnons have a unique polarization such that only in the configuration where the two ferromagnets are parallel magnons from the one can enter the other.
For antiparallel orientation, they are reflected at the second interface (see Supplemental Material for details\cite{Supplemental}).
This effect alone can produce a purely magnonic spin valve. 

For the simple models simulated, we usually find a larger spin current transmissivity for parallel alignment of the ferromagnets.
In our experiments, however, we observe a stronger spin current transport signal amplitude for the antiparallel orientation of the two ferromagnets.
While we cannot model the real system due to the complexity of YIG and the unknown YIG$|$CoO interface to check potential impacts of the complex spin current spectrum and interface transmissivity, we explore in addition to the magnonic effect in the YIG/CoO bilayer also electronic effects such as a spin-dependent ISHE in the Co layer or a spin-dependent effective spin transmissivity at the CoO$|$Co interface.
The magnonic spin current flowing across the CoO-layer generates a spin accumulation in the Co-layer that results in a pure spin current flow of spin-up and spin-down electrons moving in opposite directions.
Due to the ISHE electrons of both spin polarization are scattered towards the same direction, generating an electrical voltage measured in the experiments.
Non-magnetic metals like Pt are spin-unpolarized and regarding their properties electrons of both polarizations are equal.
A sign change of the spin current realized by switching the YIG magnetization yields a sign change of the electrical voltage, but the amplitudes are similar, as shown in YIG$|$CoO$|$Pt multilayers by Qiu et al.\cite{Qiu2016}.
Ferromagnetic metals like Co, however, are spin-polarized and exhibit significant differences in the intrinsic properties of spin-up and spin-down electrons.
The measured ISHE voltage thus results from magnons coupling to two distinct electron channels, which can be described by a two-fluid model in the case of negligible spin-intermixing processes.
Eventually, reversing the relative alignment of the YIG and Co magnetization manifests in a reversal of the spin current polarization or likewise an interchange of the roles of spin-up and spin-down electrons in the Co.
\\
Electronic effects thus may explain the observation of asymmetric spin current transmission voltages for the different alignments between YIG and Co with $V_{\mathrm{sc}}^{\antipara} > V_{\mathrm{sc}}^{\para}$.
Since the intrinsic properties of spin-up and spin-down electrons are material specific, higher voltages for the parallel alignment of the ferromagnets could also be observed.
Note that our observations for magnonic spin currents are distinct from previous work, where when investigating thermal spin current propagation in YIG$|$Cu$|$Co, which includes a non-magnetic metal as spin-conduit and thus electronic instead of magnonic spin injection in the Co layer no significant change in the signal was reported \cite{Tian2016}.

In conclusion, we demonstrated the magnon spin valve effect in YIG$|$CoO$|$Co multilayers by showing that the spin current transmission signal amplitudes depend on the relative alignment of YIG and Co.
The total voltage signal measured includes two contributions whose signs depend individually on the YIG and Co magnetization directions and thus yield different signal amplitudes and signs.
This enables one to encode even two bits of information in the magnetic configuration of the spin valve.
The presented setup gives a new insight into spin-dependent transport effects in ferromagnetic metals and provides a missing switch component for magnon-based applications thus making the work a key step towards further magnon-based logic gate operation.

\clearpage

\section*{Acknowledgments}
This work was supported by the Deutsche Forschungsgemeinschaft (DFG) (SPP 1538 “Spin Caloric Transport", SFB767 in Konstanz), the Graduate School of Excellence Materials Science in Mainz (MAINZ), the EU projects (IFOX NMP3-LA-2012246102, INSPIN FP7-ICT-2013-X 612759), ERATO "Spin Quantum Rectification Project" (No. JPMJER1402) from JST, Japan, Grant-in-Aid for Scientific Research on Innovative Area "Nano Spin Conversion Science" (No. JP26103005) and Grant-in-Aid for young scientists (B) (No. JP17K14331) from JSPS KAKENHI, Japan. 
We thank the DAAD SpinNet and MaHoJeRo projects for supporting the Tohoku-Mainz collaboration and M. K. thanks ICC-IMR at Tohoku University for their hospitality during a visiting researcher stay at the Institute for Materials Research.
T.K. is supported by JSPS through a research fellowship for young scientists (No. JP15J08026).

\section*{Methods}

\textbf{\sffamily Sample preparation} Single crystalline YIG ($d \approx \SI{5}{\micro\meter}$) was grown by liquid phase epitaxy and cut into samples of size $\SI{2}{\milli\meter} \times \SI{3}{\milli\meter}$ to ensure equal bulk properties.
To promote the growth of the CoO layer, the YIG surface was optimized by means of a rapid thermal annealing process.
In an infrared furnace, YIG samples were pairwise arrange face-to-face and heated up to \SI{1173}{\kelvin} at a heating rate of \SI{50}{\kelvin\per\second}.
The final temperature was kept for \SI{30}{\minute}, before the samples were cooled down quickly to room temperature again.
As a result, terrace like, smooth YIG surfaces of $\approx \SI{1.11}{\angstrom}$ roughness were obtained, see Supplemental Information\cite{Supplemental}.
Subsequent to the annealing procedure, CoO was deposited by reactive magnetron sputtering from a Co target in an Ar/O atmosphere employing a QAM-4-STSCP (ULVAC Inc.) system.
During the deposition process, the substrate temperature was kept at \SI{723}{\kelvin}.
X-Ray spectroscopy revealed a CoO growth along the [111] direction on top of annealed YIG samples.
Eventually, Co top layers were grown in-situ at room temperature by non-reactive sputtering from the same Co target.
Here, thickness-combinations of $d_{\mathrm{CoO}}/d_{\mathrm{Co}} = \SI{2}{\nano\meter}/\SI{4}{\nano\meter}, \SI{3}{\nano\meter}/\SI{4}{\nano\meter}$, and $\SI{5}{\nano\meter}/\SI{6}{\nano\meter}$ were realized.

\textbf{\sffamily Experimental setup} Thermoelectric as well as FMR spin pumping measurements were performed in a physical property measurement (PPMS) Dynacool system, Quantum Design, Inc., which allows for temperature dependent measurements from \SIrange{10}{400}{\kelvin}.
High temperatures were avoided to prevent undesired degradation of the sample stack.
For the ANE measurements the samples were clamped in between two aluminum nitride plates with high thermal conductivity.
The top plate is heated by means of a resistive chip heater, whereas the bottom plate serves as a heat sink.
The resultant out-of-plane temperature gradient induces, in the presence of an external magnetic field, the ANE thermovoltage, which is detected by a nanovoltmeter.

For the spin pumping experiments the samples were attached to a copper-based coplanar waveguide (CPW) with the Co layer facing the CPW.
The sample fixation as well as the electrical insulation of the Co from the copper are given by a \SI{10}{\micro\meter} thick double-sided tape.
Ferromagnetic resonance of the YIG layer was achieved by feeding pulsed microwaves in the CPW while an external magnetic field is applied.
The frequency of applied microwaves is $f = \SI{4.5}{\giga\hertz}$ with a typical applied microwave power of $P = \SI{23}{dBm}$.
The respective spin pumping signal was detected using a lock-in amplifier, which was triggered by the microwave source.

\section*{Author Contributions}

M.K. and U. N. proposed the study, which was refined and supervised additionally by E. S.. 
The samples were fabricated by Z. Q., T. N., J. C. and F. F. with help of M. K..
The measurements were carried out by J. C., F. F., M. K. with the help of D. H. and R. R..
The data was analyzed by F. F. and J. C..
The theoretical work was performed by U. R. and V. G. with supervision from U. N. and J. S. 
All authors participated in the discussion and interpreted results.
J. C. drafted the manuscript with the help of
M.K. and all authors commented on the manuscript.

\bibliography{bibliography}

\end{document}